\documentclass[times,twocolumn,final]{elsarticle}

\usepackage{outline}
\usepackage{pmgraph}
\usepackage[normalem]{ulem}
\usepackage[utf8]{inputenc}
\usepackage{amssymb}
\usepackage{hyperref}
\usepackage{amsmath}
\usepackage{graphicx}
\usepackage{times}
\usepackage{xcolor}
\usepackage{xspace}
\usepackage[colorinlistoftodos]{todonotes} 
\usepackage{bm} 
\usepackage{url}

\usepackage{epstopdf}
\AppendGraphicsExtensions{.tiff}
\graphicspath{{fig/}} 

\usepackage{epsfig}
\usepackage{tikz}
\usetikzlibrary{spy}
\usepackage{algpseudocode}
\usepackage{algorithm}
\usepackage{mathrsfs}

\def\QED{~\rule[-1pt]{5pt}{5pt}\par\medskip}

\long\def\comment#1{} 

\newcommand{\xmath}[1] {\ensuremath{#1}\xspace}
\newcommand{\blmath}[1] {\xmath{\bm{#1}}}

\newcommand{\x}{\blmath{x}}

\newcommand{\beq}{\begin{equation}}
\newcommand{\eeq}{\end{equation}}
\newcommand{\beqa}{\begin{eqnarray}}
\newcommand{\eeqa}{\end{eqnarray}}

\usepackage[mathlines]{lineno}

\usepackage{multirow}
\usepackage{makecell}
\usepackage{booktabs}
\usepackage[flushleft]{threeparttable}
\usepackage{etoolbox}
\usepackage{amsmath,soul}
\AtBeginEnvironment{figure}{\setlength{\intextsep}{0.5\baselineskip}}

\usepackage{medima}
\usepackage{framed,multirow}

\usepackage{latexsym}

\usepackage{url}
\usepackage{xcolor}
\usepackage{hyperref}

\definecolor{newcolor}{rgb}{.8,.349,.1}

\begin{document}

\verso{J. Huh \textit{et~al.}}

\begin{frontmatter}
\title{Improving Medical Speech-to-Text Accuracy with Vision-Language Pre-training Model}

\author[1]{Jaeyoung Huh\corref{cor1}}
\author[1]{Sangjoon Park\corref{cor1}}
\author[2]{Jeong Eun Lee\corref{cor2}}
\author[3]{Jong Chul Ye\corref{cor2}}
\cortext[cor1]{Co-first authors}
\cortext[cor2]{Co-corresponding authors:  E-mail address : leeje290@gmail.com (J.E.Lee); jong.ye@kaist.ac.kr (J.C.Ye)}
\cortext[]{To validate the effectiveness of AI-based error correction, human authors initially wrote this paper. Subsequently, it was reviewed and edited by ChatGPT and eventually validated by human authors.}

\address[1]{Department of Bio and Brain Engineering, Korea Advanced Institute of Science and Technology (KAIST), Daejeon 34141, Republic of Korea}
\address[2]{Department of Radiology, Chungnam National University Hospital, Chungnam National University College of Medicine, 282 Munhwa-ro, Jung-gu,  Daejeon 35015, Korea}
\address[3]{Kim Jaechul Graduate School of AI, Korea Advanced Institute of Science and Technology (KAIST), Daejeon 34141, Republic of Korea}

\begin{abstract}
Automatic Speech Recognition (ASR) is a technology that converts spoken words into text, facilitating interaction between humans and machines. One of the most common applications of ASR is Speech-To-Text (STT) technology, which simplifies user workflows by transcribing spoken words into text. In the medical field, STT has the potential to significantly reduce the workload of clinicians who rely on typists to transcribe their voice recordings. However, developing an STT model for the medical domain is challenging due to the lack of sufficient speech and text datasets. To address this issue, we propose a medical-domain text correction method that modifies the output text of a general STT system using the Vision Language Pre-training (VLP) method. VLP combines textual and visual information to correct text based on image knowledge. Our extensive experiments demonstrate that the proposed method offers quantitatively and clinically significant improvements in STT performance in the medical field. We further show that multi-modal understanding of image and text information outperforms single-modal understanding using only text information.
\end{abstract}


\end{frontmatter}

 \begin{figure*}[t]
  \center
	\includegraphics[width=\textwidth]{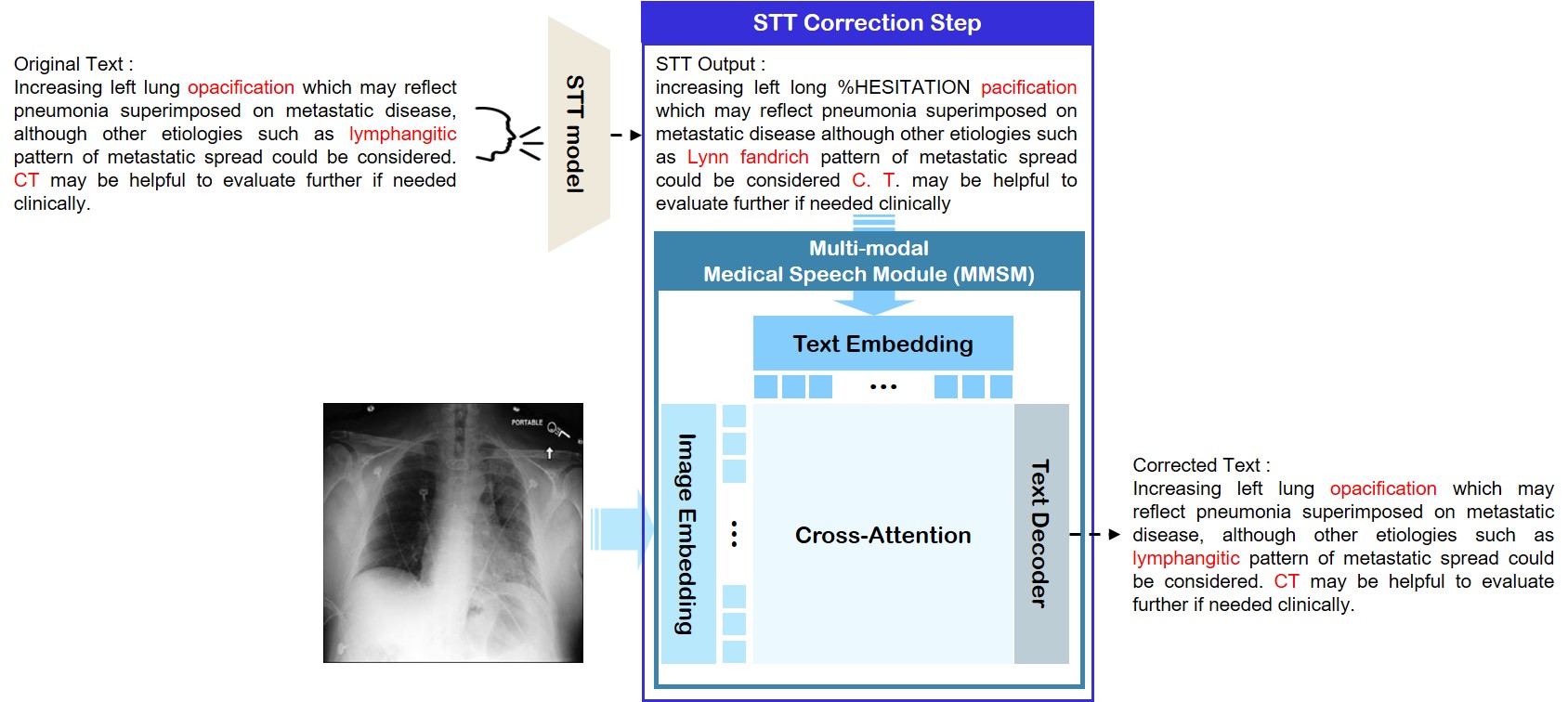}
	\caption{The proposed method consists of two main components: the Speech-To-Text (STT) system and the Multi-modal Medical Speech Module (MMSM). The STT system transcribes the speech into text, which is then input into the MMSM along with a Chest X-Ray (CXR) image. The MMSM uses both the image and the text to generate a more accurate and context-aware corrected text output.}
	\label{fig:framework}
\end{figure*}

 \begin{figure*}[t]
  \center
	\includegraphics[width=\textwidth]{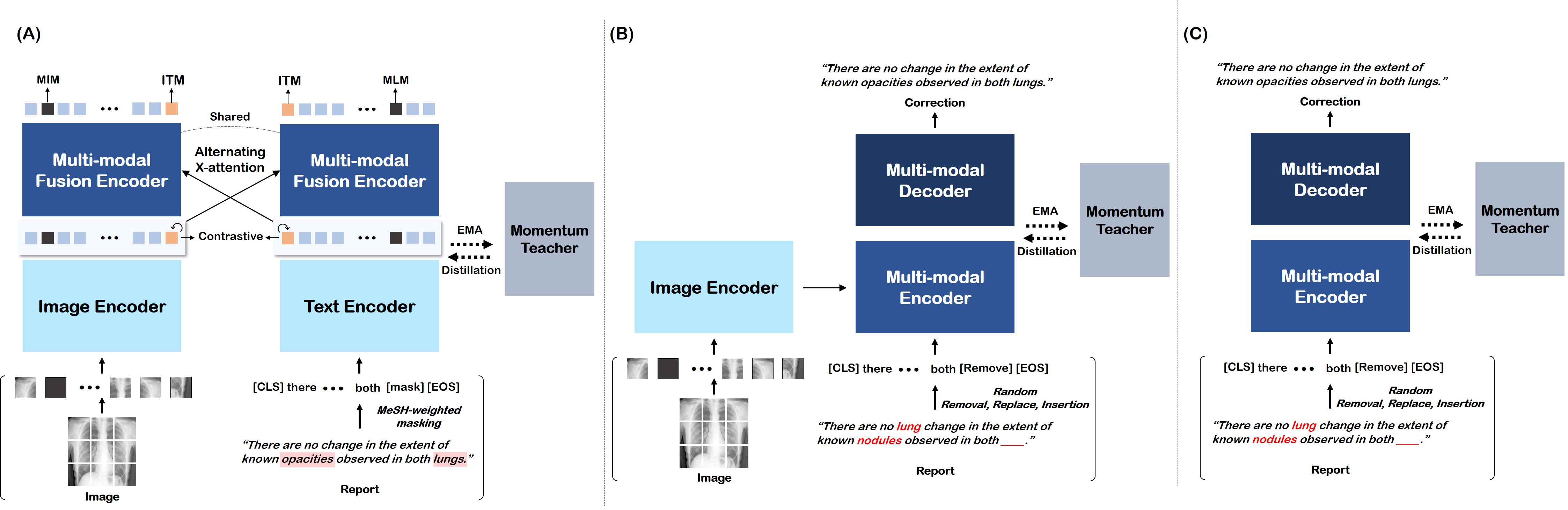}
	\caption{(A) The overall structure of the VLP model is shown, where the encoded information from the image and text encoder are fused at the multi-modal fusion encoder using alternating cross-attention.
(B) The proposed MMSM structure is shown, where the image features are fused with text features at the multi-modal encoder. The multi-modal decoder then decodes the text through a next-word prediction process, while leveraging the visual information to correct any errors.
(C) The text-only baseline model structure is shown, which is identical to the proposed MMSM model structure, except for the absence of image information fusion.}
	\label{fig:training}
\end{figure*}

\section{Introduction}
\label{sec:introduction}
Automatic Speech Recognition (ASR) is a technology that enables machines to recognize spoken language and convert it into text. By analyzing speech patterns, ASR determines which words are spoken and transcribes them into text. This technology has become widely utilized in speech-controlled devices and virtual assistants, enabling hands-free interaction and making communication more convenient. 

One of the most popular applications of ASR is the speech-to-text (STT) model, which transcribes speech into text in real-time. 
Traditionally, probabilistic models such as Hidden Markov Models (HMMs) and Gaussian Mixture Models (GMMs) have been used to transcribe speech into text (\cite{juang1991hidden,rabiner1989tutorial,reynolds2009gaussian}). 
Recent advances in deep learning frameworks have led to the development of more accurate and sophisticated STT models, which have improved the performance and usability of this technology.
For example, an open source STT model from Baidu used recurrent neural networks (RNNs) to design the sequential model, therefore, it can convert speech to text with low latency and high fidelity (\cite{hannun2014deep,amodei2016deep}). 
Facebook introduced a highly accurate and efficient STT model called wav2letter and wav2letter++ which extracts the feature of speech signal using Convolutional Neural Networks (CNNs) and designs the sequential structure through the RNNs (\cite{collobert2016wav2letter,pratap2019wav2letter++}). Google offers a cloud-based Speech-to-Text API to convert spoken language into written text in real time. Microsoft offers a cloud-based Azure STT service trained on a large dataset corpus. Aside from the aforementioned STT models by the big tech companies, there are widely used representative speech recognition tool-kits such as Kaldi and Julius, which supports a wide range of speech recognition based on probabilistic models or neural networks and language processing module (\cite{povey2011kaldi,lee2001julius}).

Medical STT applications are typically used to transcribe spoken medical dictation into written text, reducing congestion in the workflow and making it more efficient. However, achieving powerful medical STT applications is challenging due to some technical hurdles. Medical terminology and language can be complex and nuanced, and STT models may not have the knowledge required to accurately convert these terms. Therefore, STT models trained on general language may not work well in medical dictations. Moreover, STT models may need to be trained or adapted to a specific medical domain, which can be time-consuming and resource-intensive. Additionally, medical datasets may have privacy and security issues that make them difficult to access.

On the other hand, 
thanks to the advances of the self-supervised learning (SSL) in Natural Language Processing (NLP) and vision tasks, vision language pretrained (VLP) models \citep{su2019vl,lu2019vilbert,li2019visualbert,li2020unicoder,li2021align,chen2020uniter,radford2021learning}, which aims to learn the shared semantic information of vision and speech, have been studied to provides improved representations and is applied to many different applications in the medical field \citep{moon2022multi,yan2022clinical}.

Inspired by these success, we propose a Multi-modal Medical Speech Module (MMSM) that can enhance the performance of the STT model for the medical domain by correcting the output of the STT from a common language-trained model through a  pretrained VLP method (see Fig.~\ref{fig:framework}). Specifically, this method leverages multi-modal understanding of semantics to enable visual context-aware medical speech recognition by utilizing not only text concepts but also visual semantics.
Our experiments demonstrate that the proposed method outperforms the text-only Medical Speech Module (MSM), highlighting the effectiveness of leveraging visual semantics in addition to textual information.

Our contributions can be summarized as follows:
\begin{itemize}
    \item We propose a novel text correction method from various freely available STT systems to be specialized into the medical domain.
    \item Our method can be applied to any STT system trained with common language.
    \item In contrast to existing methods, the proposed method exploits the visual semantics to correct texts from STT module in a clinically meaningful way.
\end{itemize}   

This paper is structured as follows. First, we provide background information on medical VLP and STT. Next, we detail the main contribution of our proposed method in Section \ref{sec:contribution}, followed by the implementation details in Section \ref{sec:implementation}. We then present and discuss the experimental results in Section \ref{sec:results} and Section \ref{sec:Discussion}, respectively. Finally, we conclude our findings.

\section{Background}
\label{sec:background}

\subsection{Vision-language models in medical imaging}
\label{sec:vision language model}
Recent advancements in deep learning models have shown remarkable success in various tasks, but their expertise remains limited to specific domains of applications.
For instance, while computer-aided diagnosis (CAD) models based on artificial intelligence (AI) outperform experienced human readers, they lack the capacity to correlate meaningful visual semantics in medical images with keywords in medical reports, a task easily accomplished by human experts.

To address this gap in visual semantics and language understanding, VLP has recently gained popularity \citep{su2019vl,lu2019vilbert,li2019visualbert,li2020unicoder,li2021align,chen2020uniter,radford2021learning}.
 Unlike traditional learning strategies, where the model learns underlying patterns through handcrafted data-label pairs, VLP allows the model to directly learn the relationship between corresponding images and text by using uncurated image-text data corpus. This is particularly advantageous in medical imaging, where expert annotation is expensive and difficult to obtain.

Several studies have demonstrated the successful application of VLP in the medical domain \citep{park2022self,moon2022multi}, showing that VL pre-trained models achieve state-of-the-art performance in various downstream tasks, including report generation and vision-question answering. However, these applications are limited to the correlation between written text and medical images, and the use of vision-language models for speech and images has yet to be investigated.

\subsection{STT model for medical domain}

STT systems convert spoken language into written text. Conventionally, probabilistic models like HMMs and GMMs were used to transcribe speech \citep{juang1991hidden,rabiner1989tutorial,reynolds2009gaussian,gales2008application}, but they have limitations like requiring large annotated datasets and struggling with complex speech signals.


Recently, deep learning frameworks have shown promising results for STT. RNNs \citep{robinson1996use,graves2013speech,hori2018end,hannun2014deep} and DNNs \citep{hinton2012deep,dahl2011context,seide2011conversational} have been widely used, and CNNs \citep{sainath2013improvements,abdel2014convolutional,li2019jasper,zeghidour2018fully,han2020contextnet,amodei2016deep,collobert2016wav2letter,pratap2019wav2letter++} have achieved superior performance for feature extraction. Self-attention based transformer architectures have further boosted performance, with methods like two-dimensional attention \citep{dong2018speech}, combining CNN and Transformer \citep{gulati2020conformer}, and weak attention modules \citep{shi2020weak}. SSL, which trains the model to contain the language representation, offers innovative performance with fewer datasets and shorter computational resources. Pre-trained language models like ``wav2vec" and ``wav2vec2.0" \citep{schneider2019wav2vec,baevski2020wav2vec} from Facebook demonstrate high performance with simple fine-tuning using only 10 minutes of labeled data.

On the other hand, medical STT models are developed using deep learning, and they face challenges due to the complexity of medical-specific language and the lack of annotated data sets. Despite these limitations, \citep{lybarger2017automatically} proposed error detection and reformulation using logistic regression and conditional random field models. \citep{mani2020towards} attempted to correct the text from the Google ASR system using domain matching. \citep{salimbajevs2022automatic,gruzitis2022adapting} proposed language model adaptation methods. Companies such as Nuance Dragon Medical, IBM Watson, Google, and Amazon offer medical-specific STT services.
In contrast to general domain STT models, medical-specific STT models are typically proprietary solutions, making it difficult to achieve synergy with other medical domain pre-trained models like VLP.

\section{Main Contribution}
\label{sec:contribution}

\subsection{VLP-based STT Correction Model}

The vision-language pre-trained model has a thorough understanding of image and text features and their relationships, making it an ideal baseline for developing the MMSM model. The MMSM model can leverage visual semantics to enhance the accuracy of textual tasks such as speech recognition.

We utilized VLP to generate a multi-modal model with paired image-report data, using the medical X-VL \citep{park2022self} as the VL pre-trained baseline. The architecture and learning objectives of the X-VL model \citep{park2022self} are illustrated in Figure \ref{fig:training}(A). The model employs cross-attention to obtain a comprehensive representation of the image and text, using X-shaped cross-attention to process image-to-text and text-to-image fusions alternately. To increase the similarity between the image and text features from each uni-modal encoder, cross-modal contrastive (CMC) learning and intra-modal contrastive (IMC) learning were performed, having the image and text features be in the same embedding space. This enables modality-agnostic cross-attention, by alternately applying the image and text features as key/value and query for more efficient training of the model with limited data pairs.

The VLP model is optimized with five learning objectives, including masked language modeling (MLM), masked image modeling (MIM), cross-modal contrastive (CMC) loss, intra-modal contrastive (IMC) loss, and image-text matching (ITM) loss \citep{park2022self}. We used momentum distillation and hard negative mining strategies, which significantly affect the overall performance of the VLP model \citep{park2022self}. Additionally, we adopted the medical subheading (MeSH) keyword weighted masking during MLM to enhance the performance \citep{park2022self}.

The VLP network architecture used ViT-S/16 as the visual encoder, a 12-layered transformer with 12 attention heads and using the first six layer as text encoder and the last six layer as fusion encoder, respectively. We used alternative X-attention and the momentum teacher model to improve the model's vision and language representations.

We implemented a downstream MMSM model using the vision-question answer downstream model to receive image and erroneous report as input, and yield the corrected report as depicted in Figure \ref{fig:training}(B). We used the same visual encoder as VLP and the first six layers of the text encoder as the multi-modal text encoder with cross-attention. The multi-modal representations are then utilized by the multi-modal decoder to output a corrected report. Additionally, as for comparision,
we implemented a text-only speech module, as shown in Figure \ref{fig:training}(C), which uses only text information to make corrections with the uni-modal transformer encoder-decoder architecture.

\subsection{MMSM model training}

The MMSM model was fine-tuned using a pre-trained VLP model, with an auto-regressive language modeling objective. The model was trained to predict the next word in a sequence given the previous word, by minimizing the cross-entropy loss between the label and the predicted next word. The \texttt{[CLS]} token was used as the start of the sequence, and the network generated the next sequence until the \texttt{[SEP]} token was reached.

\begin{figure}[t!]
  \center
	\includegraphics[width=0.5\textwidth]{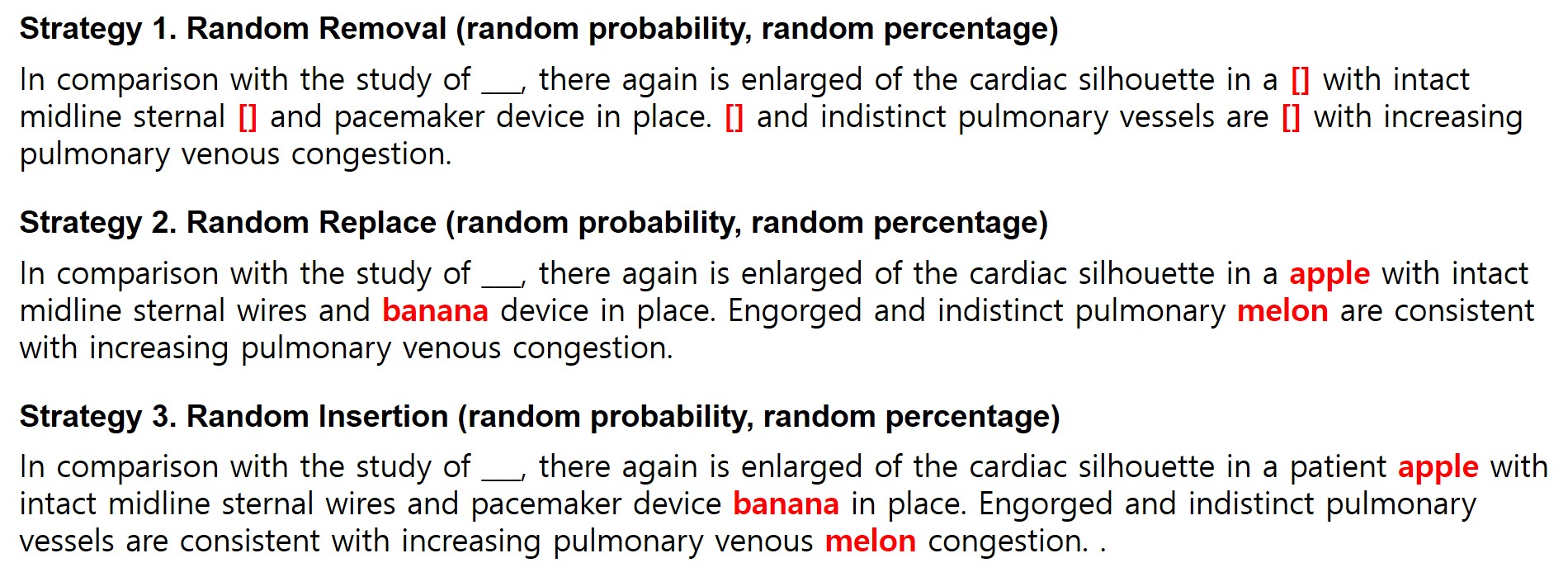}
	\caption{Our training strategy for the MMSM consists of three stages. Firstly, we employ random word removal, followed by random word replacement, and finally, we randomly insert words.}
	\label{fig:training_strategy}
\end{figure}

We further developed our own training strategy that mimicked various speech corruption cases. Specifically, we applied three strategies: randomly removing words to simulate missing speech, randomly replacing words to simulate misunderstandings by the STT model, and randomly inserting words to simulate token splitting. The process is illustrated in Fig. \ref{fig:training_strategy}. For each iteration, the probability of removal, replacement, and insertion was randomly selected between 50\% to 90\%. Additionally, we fixed the percentage of processed words in each sentence at 50\%. We selected the parameters that produced the best performance. These techniques are used to generate synthetic errors in the training dataset, which helps the MMSM learn to correct errors more effectively. By gradually increasing the complexity of the synthetic errors, we aim to enhance the robustness of the MMSM and improve its performance in correcting real-world errors.

\begin{figure*}[t!]
  \center
	\includegraphics[width=0.9\textwidth]{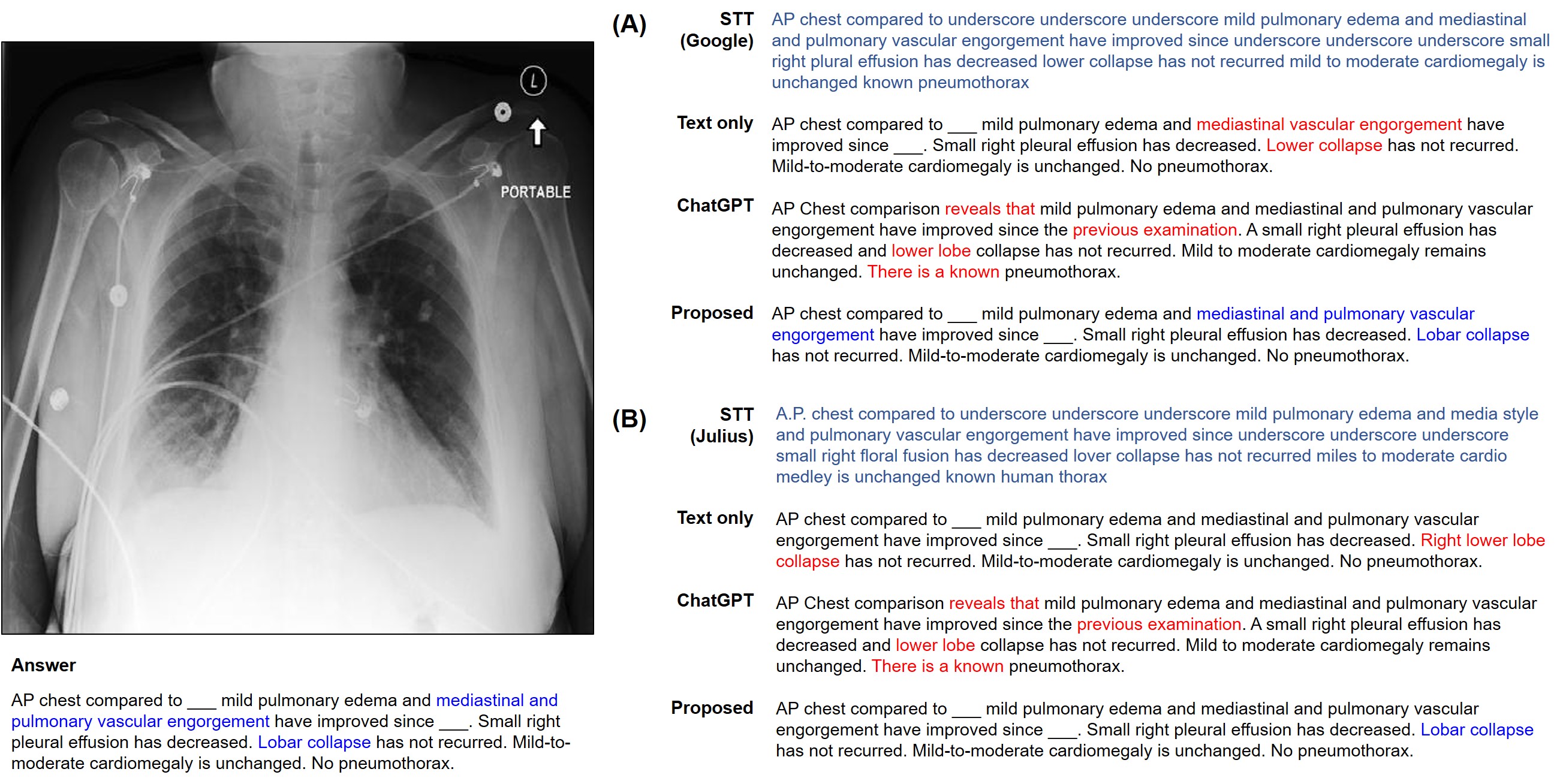}
	\vspace*{-0.3cm}
	\caption{Comparison of results using different error patterns generated from two different STT systems using the same sentence. (A) Results obtained using the Google STT system. (B) Results obtained using the Julius STT system.}
	\label{fig:result_case1}
\end{figure*}

\section{Implementation Details}
\label{sec:implementation}

\subsection{Dataset}

To train the VLP model, we utilized the MIMIC-CXR dataset, which comprises 377,110 chest X-rays from 227,827 imaging studies conducted at the Beth Israel Deaconess Medical Center between 2011 and 2016 (\cite{johnson2019mimic}). We selected anterior-posterior (AP) images and performed pre-processing based on prior literature, resulting in 91,685 images for our experiments (\cite{moon2022multi}). The MIMIC-CXR dataset was split into training, validation, and test sets, with 89,395, 759, and 1,531 images, respectively, following the official split of the MIMIC-CXR.

For the inference step, we acquired a speech dataset and ran it through various STT models. To create the speech dataset, we used the ``Balabolka" text-to-speech (TTS) program, which provides English conversion with a female voice. The default pitch and volume settings were used, resulting in 1,531 speech files generated from the test dataset mentioned above, saved as .wav audio files with a sample rate of 16,000 Hz.

%

\subsection{Details of model training}

The VLP model was pre-trained using the Adam optimizer with decoupled weight decay (AdamW), along with a warm-up cosine annealing scheduler (\cite{loshchilov2016sgdr}). The initial learning rate for the warm-up phase was set to $1e-5$, which was gradually increased to a maximum learning rate of $1e-4$ before being decayed using the cosine annealing scheduler. The visual encoder was initialized with self-supervised weights from ImageNet, while the text encoder was trained from scratch. The model was trained for 15 epochs with a warm-up epoch of 5, and a batch size of 12. All VLP model training was performed on a GeForce RTX 3090.

For fine-tuning, we used the same optimizer and scheduler, with a learning rate of $1e-4$ and a warm-up learning rate of $1e-5$. The model was initialized with the pre-trained parameters, which had already learned text-image alignment during the pre-training phase. We trained the entire network for 20 epochs with a batch size of 32 on a GeForce RTX 3090.

\begin{figure*}[t!]
  \center
	\includegraphics[width=\textwidth]{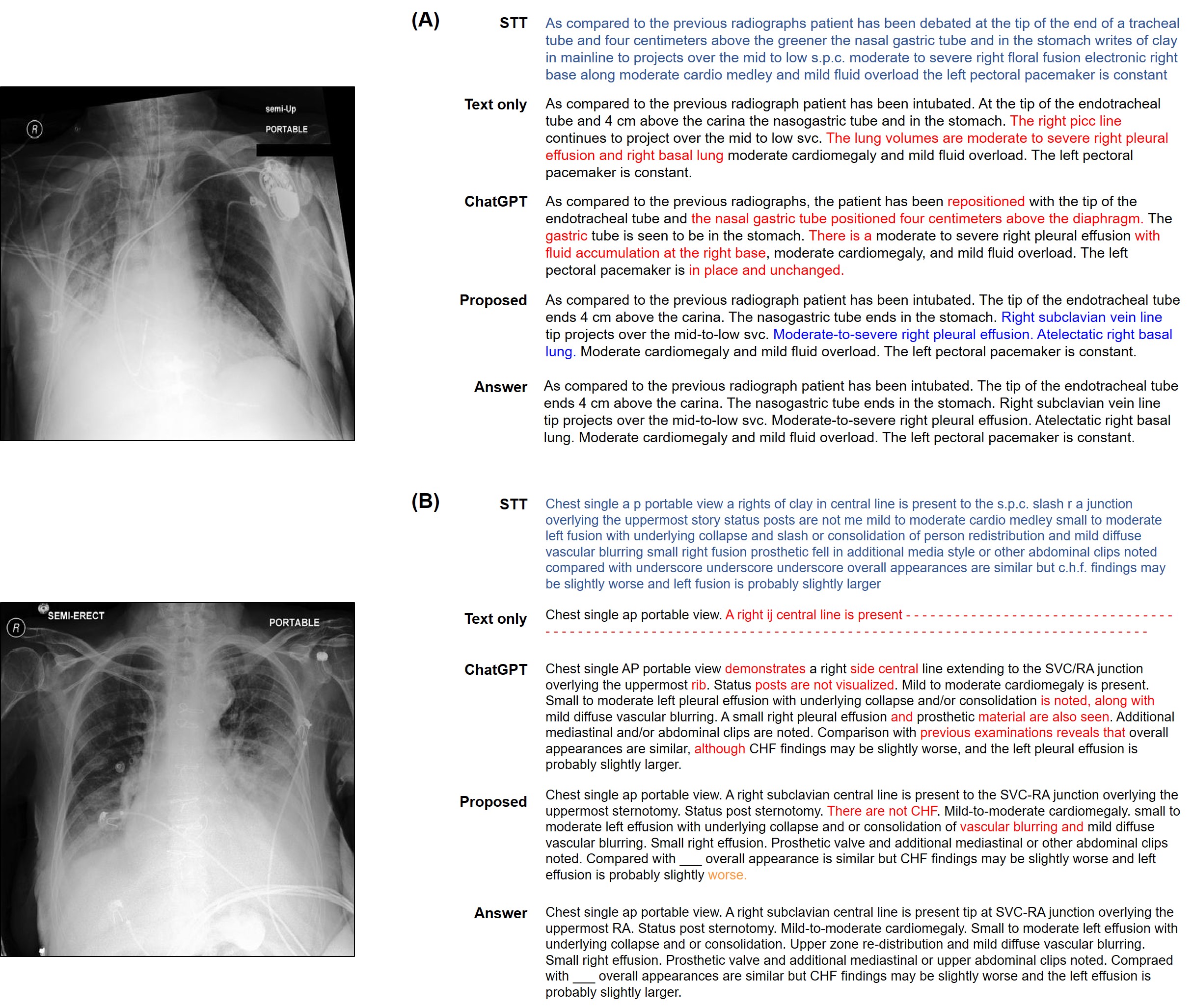}
	\vspace*{-0.3cm}
	\caption{The figures show the comparison results between the text-only model, ChatGPT, and the proposed MMSM method.
(A) The first figure shows the comparative results for clinically significant importance. The proposed MMSM method outperforms both the text-only model and ChatGPT in terms of accuracy for clinically important phrases.
(B) The second figure shows the comparison results for a long and complex sentence. Again, the proposed MMSM method performs better than both the text-only model and ChatGPT, producing a more accurate transcription of the sentence.}
	\label{fig:result_case2}
\end{figure*}

\begin{figure*}[t!]
  \center
	\includegraphics[width=\textwidth]{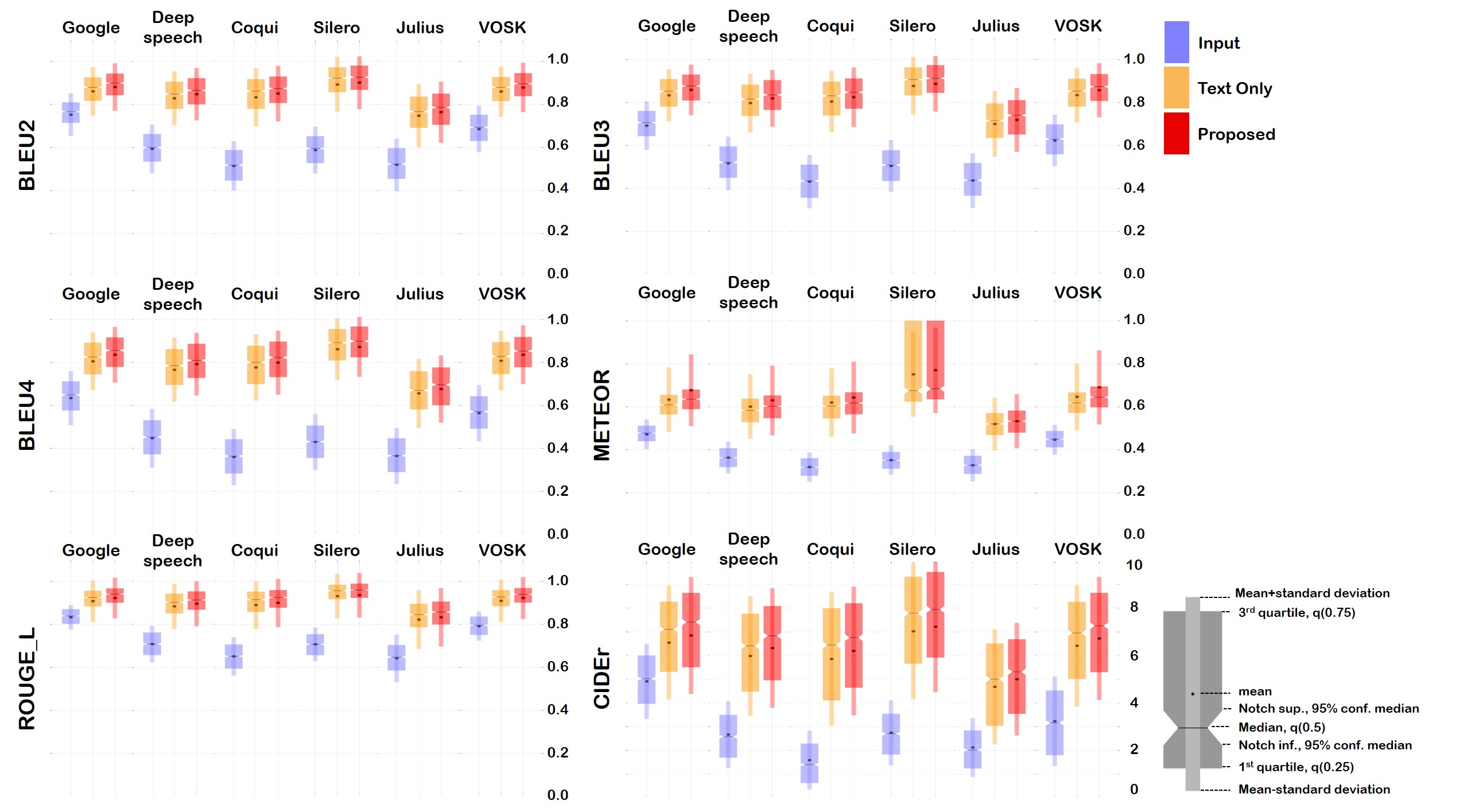}
		\vspace*{-0.3cm}
	\caption{
	Quantitative comparison results. We evaluated six different STT systems using six different evaluation metrics. In each block, each column represents the results of each STT system. The blue graph represents the input, the yellow graph represents the text-only model, and the red graph represents the proposed method results. The shape of each graph is explained in the legend shown in the lower right corner of the figure.}
	\label{fig:result_quan}
\end{figure*}

\section{Experimental Results}
\label{sec:results}

\subsection{Qualitative results}

We evaluated our method using several representative STT systems, including Google speech recognition software, Baidu's open-source STT engine DeepSpeech, the open-source deep learning toolkit Coqui, the widely-used Pytorch-based pre-trained model Silero, the well-known speech recognition toolkit Julius, and the lightweight model for portable use VOSK. We used the 1531 paired image-text test set to correct the output of each STT system using the text-only model and our proposed method, as well as ChatGPT, which has demonstrated impressive performance on various tasks. We used red and blue colors to indicate significant sets of words showing differences, with red representing wrong sentences and blue representing correct sentences compared to the answer. We used yellow to represent words that are not the same as the answer, but are not abnormal.

We selected three representative cases to illustrate the significance of the performance differences. In the first case, we observed different error patterns produced by different STT models. Figure \ref{fig:result_case1} (A) shows the correction of STT text from the Google STT system using the text-only model, ChatGPT, and our proposed method. The proposed method produced clinically significant corrections compared to the text-only model and ChatGPT when compared with the CXR image and answer. In particular, the text-only model incorrectly translated ``lobar collapse" to ``lower collapse," while the proposed method corrected it well thanks to the understanding of visual semantics. In Figure \ref{fig:result_case1} (B), we used the same sentence translated with Julius and found that the proposed method produced the same correction as in case Figure \ref{fig:result_case1} (A), although the text-only model incorrectly translated the phrase ``lobar collapse” to ``right lower lobe collapse”.
Interestingly, ChatGPT did not perform perfectly, but was able to correct the meaning to some extent in both cases.

The second case involved the correction of clinically important information. Figure \ref{fig:result_case2} (A) shows the correction of the STT output using the text-only model, ChatGPT, and our proposed method. The text-only model incorrectly translated ``writes of clay in mainline" to ``the right picc line," which is incorrect when judged against the CXR image. However, the proposed method translated it to ``right subclavian vein line," which is the same as the answer. Furthermore, the proposed method corrected the sentence ``the lung volumes are moderate to severe right pleural effusion and right bassal lung," which the text-only model failed to do. Unfortunately, ChatGPT did not perform as well in completely correcting the sentence.

The third case involved the correction of a long and complex sentence. Figure \ref{fig:result_case2} (B) shows that the text-only model was no longer able to generate a coherent answer after a certain point. Although our proposed method could not generate a fully coherent answer, it produced a significant correction compared to the STT output and the text-only model. In particular, ChatGPT corrected semantically incorrect sentences but generated sentences as a whole.

\begin{table*}[t!]
	\caption{Word Error Rate (WER) \& Character Error Rate (CER)}
    \vspace{0.0cm}
    \begin{center}
    \resizebox{0.75\textwidth}{!}{
		\begin{tabular}{c||c|c|c|c|c|c}
			\hline 
			(WER/CER) & Google & DeepSpeech & Coqui & Silero & Julius & VOSK \\ \hline \hline
            Input & 0.195/0.079 & 0.362/0.157& 0.446/0.145 & 0.350/0.140 & 0.446/0.194 & 0.273/0.105 \\ \hline
            Text-Only & 0.123/0.072 & 0.1566/0.080& 0.153/0.079 & 0.097/0.036 & 0.237/0.122 & 0.126/0.061 \\ \hline
            Proposed & \textcolor{red}{0.103/0.052} & \textcolor{red}{0.139/0.068} & \textcolor{red}{0.139/0.066} & \textcolor{red}{0.091/0.031} & \textcolor{red}{0.220/0.109} & \textcolor{red}{0.108/0.044} \\ \hline
		\end{tabular}
	}
    \end{center}
	\label{table:wer}
\end{table*}

\subsection {Quantitative Results}

\paragraph{NLP evaluation metrics}
To assess the performance of the proposed method, we measured speech recognition accuracy using 6 standard natural language processing (NLP) evaluation metrics. We obtained erroneous text from 6 representative STT software systems and utilized it as input for our method. The evaluation was conducted on a test set consisting of 1531 pairs of images and text.

We present 6 evaluation metrics. The Bilingual Evaluation Understudy (BLEU) is a widely used NLP evaluation metric (\cite{papineni2002bleu}). It calculates n-gram precision, which measures the proportion of n-grams between the generated and reference text. BLEU2, BLEU3, and BLEU4 denote the BLEU score for 2, 3, and 4 grams, respectively. The METEOR metric overcomes the limitations of the BLEU score (\cite{banerjee2005meteor}). It calculates the harmonic mean of uni-gram precision and recall, which accurately represents the relationship between generated and reference text. The Recall-Oriented Understudy for Gisting Evaluation (ROUGE) is based on n-gram recall, and we used the ROUGE-L metric to calculate recall values from the longest sequence (\cite{lin2004rouge}). The Consensus-based Image Description Evaluation (CIDEr) is another commonly used metric based on the Term Frequency-Inverse Document Frequeny (TF-IDF) (\cite{vedantam2015cider}).

The results are presented in Fig. \ref{fig:result_quan}. In each graph, we plotted the results of Google, Deepspeech, Coqui, Silero, Julius, and VOSK one after the other. The blue graph represents the input, which denotes the output of the STT model. The yellow graph represents the results of the text-only model, and the red graph represents the results of the proposed method. As shown in each graph, the proposed method outperformed the input values significantly. In particular, the proposed method performed better than the text-only model in all cases. The overall distribution of the proposed method was higher than that of the input and text-only model. Moreover, the mean value of the proposed method was higher than the other results, as indicated by the star symbol.

\paragraph{Word Error Rate (WER) \& Character Error Rate (CER)}

The WER measures the proportion of correct words compared to the reference text, while the CER evaluates correct characters. Both metrics are commonly used to quantify speech recognition performance. In this study, we obtained WER and CER values and compared them with those of the input, text-only model, and the proposed method. We collected data from a test set of 1531 paired images and text and calculated average scores, which are presented in Table \ref{table:wer}. The best score among the input, text-only model, and proposed method is highlighted in red. The proposed method exhibits the best performance across all results, particularly for ``Coqui", where WER and CER show significant improvements of approximately 0.307 and 0.079, respectively.

\begin{figure*}[t!]
  \center
	\includegraphics[width=0.9\textwidth]{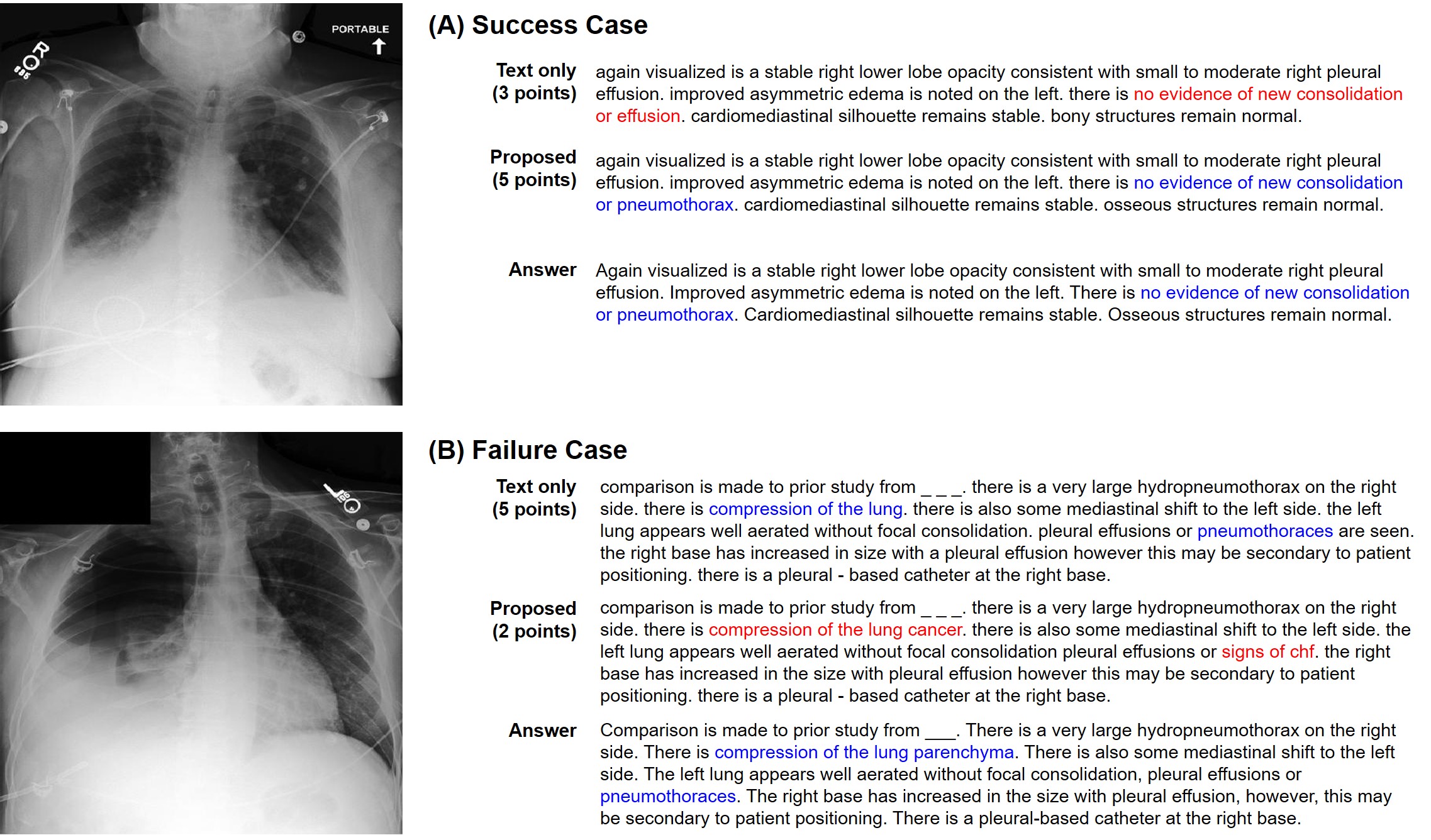}
		\vspace*{-0.3cm}
	\caption{Examples for clinical evaluation comparison. Correct corrections are denoted by blue and incorrect corrections are denoted by red compared to the reference answer. (A) A successful case where the text-only method received 3 points and the proposed method received 5 points. (B) A failure case where the text-only method received 5 points and the proposed method received 2 points.}
	\label{fig:clinical_results}
\end{figure*}

\paragraph{RadGraph F1 \& RadCliQ}

As advancements in visual-language tasks continue in the medical field, several AI-based reporting methods have been proposed to generate or create reports from medical images. However, assessing the correctness of clinical information remains a challenge, as there are no specific assessment metrics apart from natural language metrics like BLEU and METEOR. Recently, \citep{yu2022evaluating} proposed a novel metric called RadGraph F1 and RadCliQ to evaluate CXR reports, which examines the correlation between the metric and evaluated reports from radiologists. RadGraph F1 calculates relationship scores based on F1 scores related to relations and entities computed from \citep{jain2021radgraph} while RadCliQ is a composite metric that combines BLEU and RadGraph F1 using a linear regression model to predict the total number of errors. The codes supported by its authors are available at: \ul{https://drive.google.com/drive/folders/1Fe81n9IMZpc4y99K-7c5aGxPNdiij7NS?usp=sharing.}

The results are presented in Table \ref{table:rad_quan}. RadGraph F1 scores are obtained from the average between the F1 scores of relations and entities, where a higher value indicates better performance. The proposed method outperformed the text-only model in all STT systems in terms of RadGraph F1 scores. RadCliQ values that lower value indicates better performance are shown in third column at Table \ref{table:rad_quan}. All proposed methods performed better than the text-only model in terms of RadCliQ. These results demonstrate that the multi-modal method using vision and text significantly improves model performance.

\begin{table}[h!]
    \centering
    \caption{RadGraph F1 \& RadCliQ Results}
    \vspace{0.3cm}
    \resizebox{0.4\textwidth}{!}{
	\begin{tabular}{c||c|c}
			\hline 
		  (Text-Only / Proposed) & RadGraph F1 $\uparrow$ & RadCliQ $\downarrow$ \\ \hline
            Google      & 0.854/ \textcolor{red}{0.891} & 0.351/ \textcolor{red}{0.268} \\ \hline
	      DeepSpeech  & 0.833/ \textcolor{red}{0.856} & 0.438/ \textcolor{red}{0.373} \\ \hline
            Coqui       & 0.856/ \textcolor{red}{0.876} & 0.397/ \textcolor{red}{0.339} \\ \hline
            Silero      & 0.943/ \textcolor{red}{0.948} & 0.205/ \textcolor{red}{0.193} \\ \hline
            Julius      & 0.764/ \textcolor{red}{0.791} & 0.666/ \textcolor{red}{0.592} \\ \hline
            VOSK        & 0.874/ \textcolor{red}{0.907} & 0.328/ \textcolor{red}{0.253} \\ \hline
	\end{tabular}
    }
	\label{table:rad_quan}
\end{table}

\subsection {Clinical Evaluation}

The corrected text must accurately represent the corresponding image. To evaluate this, we conducted a blind assessment by a board-certified abdominal radiologist with 15 years of experience (JEL) to determine if the generated text effectively explained the image. We randomly selected 100 image-text pairs from the 1531 test set and scored them according to the criteria given in Table \ref{table:criteria}. We generated corrupted text using VOSK model and corrected it with the text-only model and the proposed method. The results are presented in Table \ref{table:clinical}. The proposed method, which combines text and image information, yielded clinically relevant results compared to the text-only model. Comparative examples are shown in Fig. \ref{fig:clinical_results} for successful and unsuccessful cases. In Fig. \ref{fig:clinical_results} (A), the text-only method received 3 points for incorrectly correcting ``no evidence of new consolidation or effusion" when the image showed a pleural effusion. In contrast, the proposed method received 5 points for correctly correcting the text. 
Although the proposed method usually generates clinical more meaningful correction than the text-only model,
it is not free of limitation.
In Fig. \ref{fig:clinical_results} (B), the proposed method received 2 points for generating incorrect sentences such as ``lung cancer" and ``signs of chf," while the text-only method received 5 points for correctly correcting the text to be clinically indistinguishable.

\begin{table}[h!]
    \centering
    \caption{Clinical Evaluation Criteria}
    \vspace{0.3cm}
	\resizebox{0.3\textwidth}{!}{
		\begin{tabular}{c|c}
			\hline 
			Score & Strength of agreement \\ \hline
			$1$ & Poor \\
			$2$ & 2 incorrect critical findings \\
			$3$ & 1 incorrect critical finding \\
			$4$ & An unimportant finding \\
			$5$ & Almost perfect \\ \hline
		\end{tabular}
	}
	\label{table:criteria}
\end{table}

\begin{table}[h!]
    \centering
    \caption{Clinical Evaluation}
    \vspace{0.3cm}
	\resizebox{0.25\textwidth}{!}{
		\begin{tabular}{c||c|c}
			\hline 
			 & Text-only & Proposed  \\ \hline
			Score &  4.35 & 4.57 \\ \hline
		\end{tabular}
	}
	\label{table:clinical}
\end{table}

\begin{figure*}[t!]
  \center
	\includegraphics[width=\textwidth]{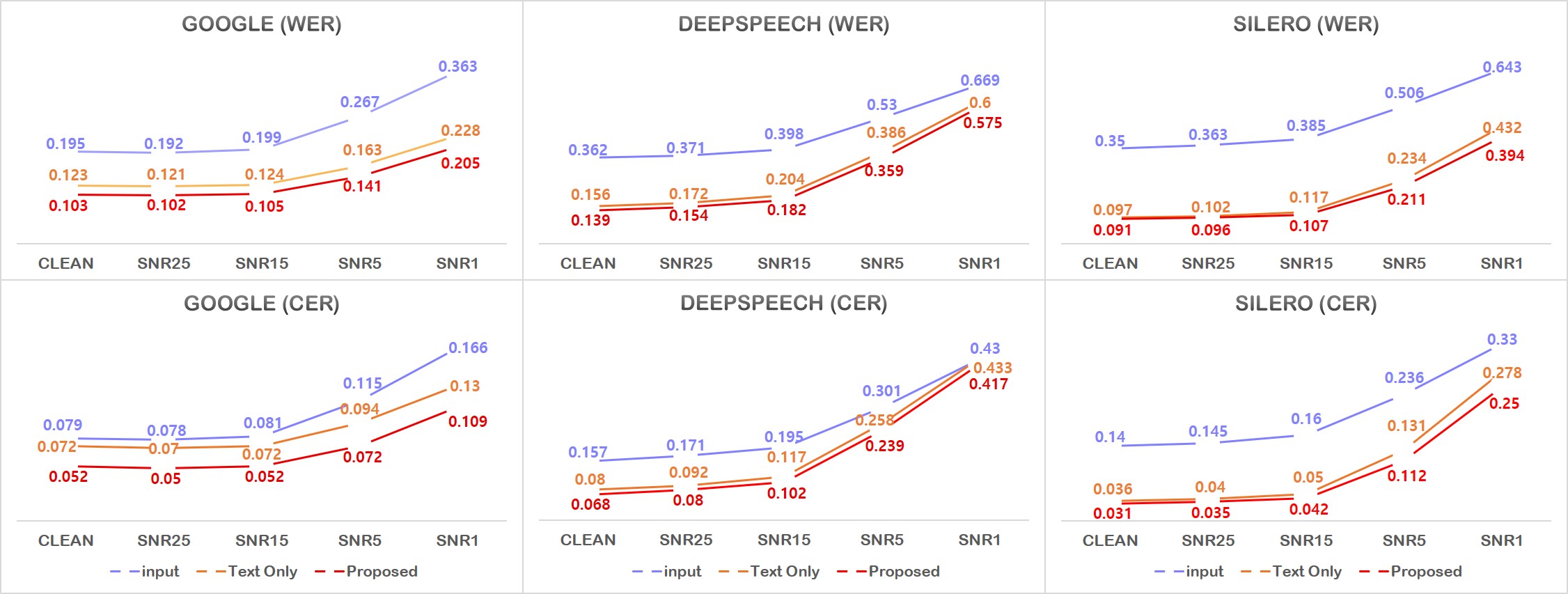}
	\vspace*{-0.3cm}
	\caption{
	Comparison results of Word Error Rate (WER) and Character Error Rate (CER) using three different STT systems at various levels of noise. Each row represents WER and CER, respectively. From left to right, each column shows the results of Google, DeepSpeech, and Silero, respectively. The blue line indicates the input, the yellow line represents the text-only model, and the red line represents the results obtained by the proposed method.
}
	\label{fig:quantitative_wer}
\end{figure*}

\subsection{Noisy Condition}

Background noise can distort speech, compromising the ability of STT systems to accurately transcribe it. To evaluate the proposed method's robustness under these conditions, we simulated them by adding white Gaussian noise at different ratios to speech signals. We degraded each signal to signal-to-noise ratios (SNRs) of 25, 15, 5, and 1 dB, with SNR1 indicating very poor quality. We then used the text-only model and the proposed method to correct the output in each case. We selected 3 STT models, Google, DeepSpeech, Silero and evaluated the performance. We calculated WER and CER values for each SNR case and STT system.

The results are shown in Figure \ref{fig:quantitative_wer}, which displays WER and CER charts for Google, DeepSpeech, and Silero, with each column representing a different STT model. The first row shows WER and the second shows CER. In each block, the values for clean, SNR25, SNR15, SNR5, and SNR1 are displayed from left to right. The results obtained using the proposed method are highlighted in red and show superior performance compared to the input and text-only methods in all cases.

\subsection{Real Speech case}

When using the STT model in real speech conditions, it is important to consider factors such as microphone quality and background noise which can significantly affect the recognition performance. To test the robustness of the proposed method under such conditions, we recorded 50 image-text pairs from the 1531 test set using mobile phones, resulting in severe corruption of the original voice signal. We then processed the corrupted voice signal with the Google STT model and evaluated the output using four assessment metrics (BLEU4, METEOR, ROUGE-L, and CIDEr), the results of which are presented in Table \ref{table:real}.

Our findings demonstrate that the proposed method is effective in correcting the corrupted STT output generated by the Google STT model, with increases of 0.242, 0.173, 0.138, and 2.404 observed in each metric respectively. In particular, the proposed method outperforms the text-only method with increases of 0.015, 0.012, 0.008, and 0.224 in each metric, respectively. The best values for each category are highlighted in red. These results highlight the robustness of the proposed method in handling real-world speech conditions.

\begin{table}[h!]
    \centering
    \caption{Evaluation Results with real speech datasets}
    \vspace{0.3cm}
    \resizebox{0.4\textwidth}{!}{
	\begin{tabular}{c||c|c|c|c}
			\hline 
			 &   BLEU4 & METEOR & ROUGE-L & CIDEr \\ \hline
		  STT(Google)     & 0.559 & 0.428 & 0.781 & 4.592 \\ \hline
	      Text Only       & 0.786 & 0.589 & 0.911 & 6.784 \\ \hline
            Proposed        & \textcolor{red}{0.801} & \textcolor{red}{0.601} & \textcolor{red}{0.919} & \textcolor{red}{7.008} \\ \hline
		\end{tabular}
  }
    \label{table:real}
\end{table}

\section{Discussion and Conclusions}
\label{sec:Discussion}

AI has found many successful applications in the medical field, and automated speech recognition is one of the most in-demand applications in radiology. Traditionally, radiologists read out the clinical findings and impressions of medical images, and typists transcribe the recordings to facilitate efficient reading processes in radiology. However, deep learning-based dictation models have recently caught up to human performance, making automated speech recognition a viable alternative.

Traditional probability-based models like HMMs-GMMs have limitations in performance and can be affected by speech conditions such as noise. While RNN, CNN, and Transformer have shown significant performance improvements, their application in clinical STT systems is not straightforward due to the need for a large amount of speech and text paired dataset, especially for medical-specific domains.

To address these limitations, we propose a novel MMSM model that leverages visual and textual information to provide a comprehensive multi-modal context-aware error correction module for STT models. Our model has a comprehensive understanding of visual semantics and textual concepts and can correct errors in the STT model by referring to both the erroneous text and the corresponding images, similar to a human expert. Our experiments show that our method consistently outperforms text-only baselines with various STT models, using both real speech and synthesized voice data.

However, our study has several limitations. First, we used generated errors to train and validate our model, as obtaining real error patterns from all STT models is not feasible. Second, we did not investigate the variety in pronunciation and voice tone, which should be considered for practical implementation. Finally, our experiments were only conducted on the CXR database, and further research is needed to validate the efficacy of our proposed MMSM for more complex medical imaging modalities like CT or MRI.

In summary, we have introduced a new method called MMSM that can effectively correct errors in STT models by leveraging comprehensive semantic context information. Importantly, this method is independent of the STT model being used, meaning that it can be applied to any STT model without requiring any additional tuning. Given that medical imaging typically involves structured pairs of medical images and reports generated from radiologists' recorded voices, our method has significant potential for application in the radiology domain.

\section*{Acknowledgement}
This work was supported by the National Research Foundation of Korea under Grant NRF-2020R1A2B5B03001980.

\bibliographystyle{model2-names.bst}\biboptions{authoryear}
\bibliography{ref}

\begin{thebibliography}{48}
\expandafter\ifx\csname natexlab\endcsname\relax\def\natexlab#1{#1}\fi
\providecommand{\url}[1]{\texttt{#1}}
\providecommand{\href}[2]{#2}
\providecommand{\path}[1]{#1}
\providecommand{\DOIprefix}{doi:}
\providecommand{\ArXivprefix}{arXiv:}
\providecommand{\URLprefix}{URL: }
\providecommand{\Pubmedprefix}{pmid:}
\providecommand{\doi}[1]{\href{http://dx.doi.org/#1}{\path{#1}}}
\providecommand{\Pubmed}[1]{\href{pmid:#1}{\path{#1}}}
\providecommand{\bibinfo}[2]{#2}
\ifx\xfnm\relax \def\xfnm[#1]{\unskip,\space#1}\fi
\bibitem[{Abdel-Hamid et~al.(2014)Abdel-Hamid, Mohamed, Jiang, Deng, Penn and
  Yu}]{abdel2014convolutional}
\bibinfo{author}{Abdel-Hamid, O.}, \bibinfo{author}{Mohamed, A.r.},
  \bibinfo{author}{Jiang, H.}, \bibinfo{author}{Deng, L.},
  \bibinfo{author}{Penn, G.}, \bibinfo{author}{Yu, D.}, \bibinfo{year}{2014}.
\newblock \bibinfo{title}{Convolutional neural networks for speech
  recognition}.
\newblock \bibinfo{journal}{IEEE/ACM Transactions on audio, speech, and
  language processing} \bibinfo{volume}{22}, \bibinfo{pages}{1533--1545}.
\bibitem[{Amodei et~al.(2016)Amodei, Ananthanarayanan, Anubhai, Bai,
  Battenberg, Case, Casper, Catanzaro, Cheng, Chen et~al.}]{amodei2016deep}
\bibinfo{author}{Amodei, D.}, \bibinfo{author}{Ananthanarayanan, S.},
  \bibinfo{author}{Anubhai, R.}, \bibinfo{author}{Bai, J.},
  \bibinfo{author}{Battenberg, E.}, \bibinfo{author}{Case, C.},
  \bibinfo{author}{Casper, J.}, \bibinfo{author}{Catanzaro, B.},
  \bibinfo{author}{Cheng, Q.}, \bibinfo{author}{Chen, G.}, et~al.,
  \bibinfo{year}{2016}.
\newblock \bibinfo{title}{Deep speech 2: End-to-end speech recognition in
  english and mandarin}, in: \bibinfo{booktitle}{International conference on
  machine learning}, \bibinfo{organization}{PMLR}. pp.
  \bibinfo{pages}{173--182}.
\bibitem[{Baevski et~al.(2020)Baevski, Zhou, Mohamed and
  Auli}]{baevski2020wav2vec}
\bibinfo{author}{Baevski, A.}, \bibinfo{author}{Zhou, Y.},
  \bibinfo{author}{Mohamed, A.}, \bibinfo{author}{Auli, M.},
  \bibinfo{year}{2020}.
\newblock \bibinfo{title}{wav2vec 2.0: A framework for self-supervised learning
  of speech representations}.
\newblock \bibinfo{journal}{Advances in neural information processing systems}
  \bibinfo{volume}{33}, \bibinfo{pages}{12449--12460}.
\bibitem[{Banerjee and Lavie(2005)}]{banerjee2005meteor}
\bibinfo{author}{Banerjee, S.}, \bibinfo{author}{Lavie, A.},
  \bibinfo{year}{2005}.
\newblock \bibinfo{title}{Meteor: An automatic metric for mt evaluation with
  improved correlation with human judgments}, in:
  \bibinfo{booktitle}{Proceedings of the acl workshop on intrinsic and
  extrinsic evaluation measures for machine translation and/or summarization},
  pp. \bibinfo{pages}{65--72}.
\bibitem[{Chen et~al.(2020)Chen, Li, Yu, El~Kholy, Ahmed, Gan, Cheng and
  Liu}]{chen2020uniter}
\bibinfo{author}{Chen, Y.C.}, \bibinfo{author}{Li, L.}, \bibinfo{author}{Yu,
  L.}, \bibinfo{author}{El~Kholy, A.}, \bibinfo{author}{Ahmed, F.},
  \bibinfo{author}{Gan, Z.}, \bibinfo{author}{Cheng, Y.}, \bibinfo{author}{Liu,
  J.}, \bibinfo{year}{2020}.
\newblock \bibinfo{title}{Uniter: Universal image-text representation
  learning}, in: \bibinfo{booktitle}{Computer Vision--ECCV 2020: 16th European
  Conference, Glasgow, UK, August 23--28, 2020, Proceedings, Part XXX},
  \bibinfo{organization}{Springer}. pp. \bibinfo{pages}{104--120}.
\bibitem[{Collobert et~al.(2016)Collobert, Puhrsch and
  Synnaeve}]{collobert2016wav2letter}
\bibinfo{author}{Collobert, R.}, \bibinfo{author}{Puhrsch, C.},
  \bibinfo{author}{Synnaeve, G.}, \bibinfo{year}{2016}.
\newblock \bibinfo{title}{Wav2letter: an end-to-end convnet-based speech
  recognition system}.
\newblock \bibinfo{journal}{arXiv preprint arXiv:1609.03193} .
\bibitem[{Dahl et~al.(2011)Dahl, Yu, Deng and Acero}]{dahl2011context}
\bibinfo{author}{Dahl, G.E.}, \bibinfo{author}{Yu, D.}, \bibinfo{author}{Deng,
  L.}, \bibinfo{author}{Acero, A.}, \bibinfo{year}{2011}.
\newblock \bibinfo{title}{Context-dependent pre-trained deep neural networks
  for large-vocabulary speech recognition}.
\newblock \bibinfo{journal}{IEEE Transactions on audio, speech, and language
  processing} \bibinfo{volume}{20}, \bibinfo{pages}{30--42}.
\bibitem[{Dong et~al.(2018)Dong, Xu and Xu}]{dong2018speech}
\bibinfo{author}{Dong, L.}, \bibinfo{author}{Xu, S.}, \bibinfo{author}{Xu, B.},
  \bibinfo{year}{2018}.
\newblock \bibinfo{title}{Speech-transformer: a no-recurrence
  sequence-to-sequence model for speech recognition}, in:
  \bibinfo{booktitle}{2018 IEEE international conference on acoustics, speech
  and signal processing (ICASSP)}, \bibinfo{organization}{IEEE}. pp.
  \bibinfo{pages}{5884--5888}.
\bibitem[{Gales et~al.(2008)Gales, Young et~al.}]{gales2008application}
\bibinfo{author}{Gales, M.}, \bibinfo{author}{Young, S.}, et~al.,
  \bibinfo{year}{2008}.
\newblock \bibinfo{title}{The application of hidden markov models in speech
  recognition}.
\newblock \bibinfo{journal}{Foundations and Trends{\textregistered} in Signal
  Processing} \bibinfo{volume}{1}, \bibinfo{pages}{195--304}.
\bibitem[{Graves et~al.(2013)Graves, Mohamed and Hinton}]{graves2013speech}
\bibinfo{author}{Graves, A.}, \bibinfo{author}{Mohamed, A.r.},
  \bibinfo{author}{Hinton, G.}, \bibinfo{year}{2013}.
\newblock \bibinfo{title}{Speech recognition with deep recurrent neural
  networks}, in: \bibinfo{booktitle}{2013 IEEE international conference on
  acoustics, speech and signal processing}, \bibinfo{organization}{Ieee}. pp.
  \bibinfo{pages}{6645--6649}.
\bibitem[{Gruzitis et~al.(2022)Gruzitis, Dargis, Lasmanis, Garkaje and
  Gosko}]{gruzitis2022adapting}
\bibinfo{author}{Gruzitis, N.}, \bibinfo{author}{Dargis, R.},
  \bibinfo{author}{Lasmanis, V.J.}, \bibinfo{author}{Garkaje, G.},
  \bibinfo{author}{Gosko, D.}, \bibinfo{year}{2022}.
\newblock \bibinfo{title}{Adapting automatic speech recognition to the
  radiology domain for a less-resourced language: the case of latvian}, in:
  \bibinfo{booktitle}{Intelligent Sustainable Systems: Selected Papers of
  WorldS4 2021, Volume 1}, \bibinfo{organization}{Springer}. pp.
  \bibinfo{pages}{267--276}.
\bibitem[{Gulati et~al.(2020)Gulati, Qin, Chiu, Parmar, Zhang, Yu, Han, Wang,
  Zhang, Wu et~al.}]{gulati2020conformer}
\bibinfo{author}{Gulati, A.}, \bibinfo{author}{Qin, J.}, \bibinfo{author}{Chiu,
  C.C.}, \bibinfo{author}{Parmar, N.}, \bibinfo{author}{Zhang, Y.},
  \bibinfo{author}{Yu, J.}, \bibinfo{author}{Han, W.}, \bibinfo{author}{Wang,
  S.}, \bibinfo{author}{Zhang, Z.}, \bibinfo{author}{Wu, Y.}, et~al.,
  \bibinfo{year}{2020}.
\newblock \bibinfo{title}{Conformer: Convolution-augmented transformer for
  speech recognition}.
\newblock \bibinfo{journal}{arXiv preprint arXiv:2005.08100} .
\bibitem[{Han et~al.(2020)Han, Zhang, Zhang, Yu, Chiu, Qin, Gulati, Pang and
  Wu}]{han2020contextnet}
\bibinfo{author}{Han, W.}, \bibinfo{author}{Zhang, Z.}, \bibinfo{author}{Zhang,
  Y.}, \bibinfo{author}{Yu, J.}, \bibinfo{author}{Chiu, C.C.},
  \bibinfo{author}{Qin, J.}, \bibinfo{author}{Gulati, A.},
  \bibinfo{author}{Pang, R.}, \bibinfo{author}{Wu, Y.}, \bibinfo{year}{2020}.
\newblock \bibinfo{title}{Contextnet: Improving convolutional neural networks
  for automatic speech recognition with global context}.
\newblock \bibinfo{journal}{arXiv preprint arXiv:2005.03191} .
\bibitem[{Hannun et~al.(2014)Hannun, Case, Casper, Catanzaro, Diamos, Elsen,
  Prenger, Satheesh, Sengupta, Coates et~al.}]{hannun2014deep}
\bibinfo{author}{Hannun, A.}, \bibinfo{author}{Case, C.},
  \bibinfo{author}{Casper, J.}, \bibinfo{author}{Catanzaro, B.},
  \bibinfo{author}{Diamos, G.}, \bibinfo{author}{Elsen, E.},
  \bibinfo{author}{Prenger, R.}, \bibinfo{author}{Satheesh, S.},
  \bibinfo{author}{Sengupta, S.}, \bibinfo{author}{Coates, A.}, et~al.,
  \bibinfo{year}{2014}.
\newblock \bibinfo{title}{Deep speech: Scaling up end-to-end speech
  recognition}.
\newblock \bibinfo{journal}{arXiv preprint arXiv:1412.5567} .
\bibitem[{Hinton et~al.(2012)Hinton, Deng, Yu, Dahl, Mohamed, Jaitly, Senior,
  Vanhoucke, Nguyen, Sainath et~al.}]{hinton2012deep}
\bibinfo{author}{Hinton, G.}, \bibinfo{author}{Deng, L.}, \bibinfo{author}{Yu,
  D.}, \bibinfo{author}{Dahl, G.E.}, \bibinfo{author}{Mohamed, A.r.},
  \bibinfo{author}{Jaitly, N.}, \bibinfo{author}{Senior, A.},
  \bibinfo{author}{Vanhoucke, V.}, \bibinfo{author}{Nguyen, P.},
  \bibinfo{author}{Sainath, T.N.}, et~al., \bibinfo{year}{2012}.
\newblock \bibinfo{title}{Deep neural networks for acoustic modeling in speech
  recognition: The shared views of four research groups}.
\newblock \bibinfo{journal}{IEEE Signal processing magazine}
  \bibinfo{volume}{29}, \bibinfo{pages}{82--97}.
\bibitem[{Hori et~al.(2018)Hori, Cho and Watanabe}]{hori2018end}
\bibinfo{author}{Hori, T.}, \bibinfo{author}{Cho, J.},
  \bibinfo{author}{Watanabe, S.}, \bibinfo{year}{2018}.
\newblock \bibinfo{title}{End-to-end speech recognition with word-based rnn
  language models}, in: \bibinfo{booktitle}{2018 IEEE Spoken Language
  Technology Workshop (SLT)}, \bibinfo{organization}{IEEE}. pp.
  \bibinfo{pages}{389--396}.
\bibitem[{Jain et~al.(2021)Jain, Agrawal, Saporta, Truong, Duong, Bui, Chambon,
  Zhang, Lungren, Ng et~al.}]{jain2021radgraph}
\bibinfo{author}{Jain, S.}, \bibinfo{author}{Agrawal, A.},
  \bibinfo{author}{Saporta, A.}, \bibinfo{author}{Truong, S.Q.},
  \bibinfo{author}{Duong, D.N.}, \bibinfo{author}{Bui, T.},
  \bibinfo{author}{Chambon, P.}, \bibinfo{author}{Zhang, Y.},
  \bibinfo{author}{Lungren, M.P.}, \bibinfo{author}{Ng, A.Y.}, et~al.,
  \bibinfo{year}{2021}.
\newblock \bibinfo{title}{Radgraph: Extracting clinical entities and relations
  from radiology reports}.
\newblock \bibinfo{journal}{arXiv preprint arXiv:2106.14463} .
\bibitem[{Johnson et~al.(2019)Johnson, Pollard, Greenbaum, Lungren, Deng, Peng,
  Lu, Mark, Berkowitz and Horng}]{johnson2019mimic}
\bibinfo{author}{Johnson, A.E.}, \bibinfo{author}{Pollard, T.J.},
  \bibinfo{author}{Greenbaum, N.R.}, \bibinfo{author}{Lungren, M.P.},
  \bibinfo{author}{Deng, C.y.}, \bibinfo{author}{Peng, Y.},
  \bibinfo{author}{Lu, Z.}, \bibinfo{author}{Mark, R.G.},
  \bibinfo{author}{Berkowitz, S.J.}, \bibinfo{author}{Horng, S.},
  \bibinfo{year}{2019}.
\newblock \bibinfo{title}{Mimic-cxr-jpg, a large publicly available database of
  labeled chest radiographs}.
\newblock \bibinfo{journal}{arXiv preprint arXiv:1901.07042} .
\bibitem[{Juang and Rabiner(1991)}]{juang1991hidden}
\bibinfo{author}{Juang, B.H.}, \bibinfo{author}{Rabiner, L.R.},
  \bibinfo{year}{1991}.
\newblock \bibinfo{title}{Hidden markov models for speech recognition}.
\newblock \bibinfo{journal}{Technometrics} \bibinfo{volume}{33},
  \bibinfo{pages}{251--272}.
\bibitem[{Lee et~al.(2001)Lee, Kawahara and Shikano}]{lee2001julius}
\bibinfo{author}{Lee, A.}, \bibinfo{author}{Kawahara, T.},
  \bibinfo{author}{Shikano, K.}, \bibinfo{year}{2001}.
\newblock \bibinfo{title}{Julius---an open source real-time large vocabulary
  recognition engine} .
\bibitem[{Li et~al.(2020)Li, Duan, Fang, Gong and Jiang}]{li2020unicoder}
\bibinfo{author}{Li, G.}, \bibinfo{author}{Duan, N.}, \bibinfo{author}{Fang,
  Y.}, \bibinfo{author}{Gong, M.}, \bibinfo{author}{Jiang, D.},
  \bibinfo{year}{2020}.
\newblock \bibinfo{title}{Unicoder-vl: A universal encoder for vision and
  language by cross-modal pre-training}, in: \bibinfo{booktitle}{Proceedings of
  the AAAI Conference on Artificial Intelligence}, pp.
  \bibinfo{pages}{11336--11344}.
\bibitem[{Li et~al.(2019a)Li, Lavrukhin, Ginsburg, Leary, Kuchaiev, Cohen,
  Nguyen and Gadde}]{li2019jasper}
\bibinfo{author}{Li, J.}, \bibinfo{author}{Lavrukhin, V.},
  \bibinfo{author}{Ginsburg, B.}, \bibinfo{author}{Leary, R.},
  \bibinfo{author}{Kuchaiev, O.}, \bibinfo{author}{Cohen, J.M.},
  \bibinfo{author}{Nguyen, H.}, \bibinfo{author}{Gadde, R.T.},
  \bibinfo{year}{2019}a.
\newblock \bibinfo{title}{Jasper: An end-to-end convolutional neural acoustic
  model}.
\newblock \bibinfo{journal}{arXiv preprint arXiv:1904.03288} .
\bibitem[{Li et~al.(2021)Li, Selvaraju, Gotmare, Joty, Xiong and
  Hoi}]{li2021align}
\bibinfo{author}{Li, J.}, \bibinfo{author}{Selvaraju, R.},
  \bibinfo{author}{Gotmare, A.}, \bibinfo{author}{Joty, S.},
  \bibinfo{author}{Xiong, C.}, \bibinfo{author}{Hoi, S.C.H.},
  \bibinfo{year}{2021}.
\newblock \bibinfo{title}{Align before fuse: Vision and language representation
  learning with momentum distillation}.
\newblock \bibinfo{journal}{Advances in neural information processing systems}
  \bibinfo{volume}{34}, \bibinfo{pages}{9694--9705}.
\bibitem[{Li et~al.(2019b)Li, Yatskar, Yin, Hsieh and Chang}]{li2019visualbert}
\bibinfo{author}{Li, L.H.}, \bibinfo{author}{Yatskar, M.},
  \bibinfo{author}{Yin, D.}, \bibinfo{author}{Hsieh, C.J.},
  \bibinfo{author}{Chang, K.W.}, \bibinfo{year}{2019}b.
\newblock \bibinfo{title}{Visualbert: A simple and performant baseline for
  vision and language}.
\newblock \bibinfo{journal}{arXiv preprint arXiv:1908.03557} .
\bibitem[{Lin(2004)}]{lin2004rouge}
\bibinfo{author}{Lin, C.Y.}, \bibinfo{year}{2004}.
\newblock \bibinfo{title}{Rouge: A package for automatic evaluation of
  summaries}, in: \bibinfo{booktitle}{Text summarization branches out}, pp.
  \bibinfo{pages}{74--81}.
\bibitem[{Loshchilov and Hutter(2016)}]{loshchilov2016sgdr}
\bibinfo{author}{Loshchilov, I.}, \bibinfo{author}{Hutter, F.},
  \bibinfo{year}{2016}.
\newblock \bibinfo{title}{Sgdr: Stochastic gradient descent with warm
  restarts}.
\newblock \bibinfo{journal}{arXiv preprint arXiv:1608.03983} .
\bibitem[{Lu et~al.(2019)Lu, Batra, Parikh and Lee}]{lu2019vilbert}
\bibinfo{author}{Lu, J.}, \bibinfo{author}{Batra, D.}, \bibinfo{author}{Parikh,
  D.}, \bibinfo{author}{Lee, S.}, \bibinfo{year}{2019}.
\newblock \bibinfo{title}{Vilbert: Pretraining task-agnostic visiolinguistic
  representations for vision-and-language tasks}.
\newblock \bibinfo{journal}{Advances in neural information processing systems}
  \bibinfo{volume}{32}.
\bibitem[{Lybarger et~al.(2017)Lybarger, Ostendorf and
  Yetisgen}]{lybarger2017automatically}
\bibinfo{author}{Lybarger, K.}, \bibinfo{author}{Ostendorf, M.},
  \bibinfo{author}{Yetisgen, M.}, \bibinfo{year}{2017}.
\newblock \bibinfo{title}{Automatically detecting likely edits in clinical
  notes created using automatic speech recognition}, in:
  \bibinfo{booktitle}{AMIA Annual Symposium Proceedings},
  \bibinfo{organization}{American Medical Informatics Association}. p.
  \bibinfo{pages}{1186}.
\bibitem[{Mani et~al.(2020)Mani, Palaskar and Konam}]{mani2020towards}
\bibinfo{author}{Mani, A.}, \bibinfo{author}{Palaskar, S.},
  \bibinfo{author}{Konam, S.}, \bibinfo{year}{2020}.
\newblock \bibinfo{title}{Towards understanding asr error correction for
  medical conversations}, in: \bibinfo{booktitle}{Proceedings of the first
  workshop on natural language processing for medical conversations}, pp.
  \bibinfo{pages}{7--11}.
\bibitem[{Moon et~al.(2022)Moon, Lee, Shin, Kim and Choi}]{moon2022multi}
\bibinfo{author}{Moon, J.H.}, \bibinfo{author}{Lee, H.}, \bibinfo{author}{Shin,
  W.}, \bibinfo{author}{Kim, Y.H.}, \bibinfo{author}{Choi, E.},
  \bibinfo{year}{2022}.
\newblock \bibinfo{title}{Multi-modal understanding and generation for medical
  images and text via vision-language pre-training}.
\newblock \bibinfo{journal}{IEEE Journal of Biomedical and Health Informatics}
  \bibinfo{volume}{26}, \bibinfo{pages}{6070--6080}.
\bibitem[{Papineni et~al.(2002)Papineni, Roukos, Ward and
  Zhu}]{papineni2002bleu}
\bibinfo{author}{Papineni, K.}, \bibinfo{author}{Roukos, S.},
  \bibinfo{author}{Ward, T.}, \bibinfo{author}{Zhu, W.J.},
  \bibinfo{year}{2002}.
\newblock \bibinfo{title}{Bleu: a method for automatic evaluation of machine
  translation}, in: \bibinfo{booktitle}{Proceedings of the 40th annual meeting
  of the Association for Computational Linguistics}, pp.
  \bibinfo{pages}{311--318}.
\bibitem[{Park et~al.(2022)Park, Lee, Shin, Lee and Ye}]{park2022self}
\bibinfo{author}{Park, S.}, \bibinfo{author}{Lee, E.S.}, \bibinfo{author}{Shin,
  K.S.}, \bibinfo{author}{Lee, J.E.}, \bibinfo{author}{Ye, J.C.},
  \bibinfo{year}{2022}.
\newblock \bibinfo{title}{Self-supervised co-learning of uncurated images and
  reports enables oversight ai in radiology}.
\newblock \bibinfo{journal}{arXiv preprint arXiv:2208.05140} .
\bibitem[{Povey et~al.(2011)Povey, Ghoshal, Boulianne, Burget, Glembek, Goel,
  Hannemann, Motlicek, Qian, Schwarz et~al.}]{povey2011kaldi}
\bibinfo{author}{Povey, D.}, \bibinfo{author}{Ghoshal, A.},
  \bibinfo{author}{Boulianne, G.}, \bibinfo{author}{Burget, L.},
  \bibinfo{author}{Glembek, O.}, \bibinfo{author}{Goel, N.},
  \bibinfo{author}{Hannemann, M.}, \bibinfo{author}{Motlicek, P.},
  \bibinfo{author}{Qian, Y.}, \bibinfo{author}{Schwarz, P.}, et~al.,
  \bibinfo{year}{2011}.
\newblock \bibinfo{title}{The kaldi speech recognition toolkit}, in:
  \bibinfo{booktitle}{IEEE 2011 workshop on automatic speech recognition and
  understanding}, \bibinfo{organization}{IEEE Signal Processing Society}.
\bibitem[{Pratap et~al.(2019)Pratap, Hannun, Xu, Cai, Kahn, Synnaeve,
  Liptchinsky and Collobert}]{pratap2019wav2letter++}
\bibinfo{author}{Pratap, V.}, \bibinfo{author}{Hannun, A.},
  \bibinfo{author}{Xu, Q.}, \bibinfo{author}{Cai, J.}, \bibinfo{author}{Kahn,
  J.}, \bibinfo{author}{Synnaeve, G.}, \bibinfo{author}{Liptchinsky, V.},
  \bibinfo{author}{Collobert, R.}, \bibinfo{year}{2019}.
\newblock \bibinfo{title}{Wav2letter++: A fast open-source speech recognition
  system}, in: \bibinfo{booktitle}{ICASSP 2019-2019 IEEE International
  Conference on Acoustics, Speech and Signal Processing (ICASSP)},
  \bibinfo{organization}{IEEE}. pp. \bibinfo{pages}{6460--6464}.
\bibitem[{Rabiner(1989)}]{rabiner1989tutorial}
\bibinfo{author}{Rabiner, L.R.}, \bibinfo{year}{1989}.
\newblock \bibinfo{title}{A tutorial on hidden markov models and selected
  applications in speech recognition}.
\newblock \bibinfo{journal}{Proceedings of the IEEE} \bibinfo{volume}{77},
  \bibinfo{pages}{257--286}.
\bibitem[{Radford et~al.(2021)Radford, Kim, Hallacy, Ramesh, Goh, Agarwal,
  Sastry, Askell, Mishkin, Clark et~al.}]{radford2021learning}
\bibinfo{author}{Radford, A.}, \bibinfo{author}{Kim, J.W.},
  \bibinfo{author}{Hallacy, C.}, \bibinfo{author}{Ramesh, A.},
  \bibinfo{author}{Goh, G.}, \bibinfo{author}{Agarwal, S.},
  \bibinfo{author}{Sastry, G.}, \bibinfo{author}{Askell, A.},
  \bibinfo{author}{Mishkin, P.}, \bibinfo{author}{Clark, J.}, et~al.,
  \bibinfo{year}{2021}.
\newblock \bibinfo{title}{Learning transferable visual models from natural
  language supervision}, in: \bibinfo{booktitle}{International conference on
  machine learning}, \bibinfo{organization}{PMLR}. pp.
  \bibinfo{pages}{8748--8763}.
\bibitem[{Reynolds et~al.(2009)}]{reynolds2009gaussian}
\bibinfo{author}{Reynolds, D.A.}, et~al., \bibinfo{year}{2009}.
\newblock \bibinfo{title}{Gaussian mixture models.}
\newblock \bibinfo{journal}{Encyclopedia of biometrics} \bibinfo{volume}{741}.
\bibitem[{Robinson et~al.(1996)Robinson, Hochberg and Renals}]{robinson1996use}
\bibinfo{author}{Robinson, T.}, \bibinfo{author}{Hochberg, M.},
  \bibinfo{author}{Renals, S.}, \bibinfo{year}{1996}.
\newblock \bibinfo{title}{The use of recurrent neural networks in continuous
  speech recognition}.
\newblock \bibinfo{journal}{Automatic speech and speaker recognition} ,
  \bibinfo{pages}{233--258}.
\bibitem[{Sainath et~al.(2013)Sainath, Kingsbury, Mohamed, Dahl, Saon, Soltau,
  Beran, Aravkin and Ramabhadran}]{sainath2013improvements}
\bibinfo{author}{Sainath, T.N.}, \bibinfo{author}{Kingsbury, B.},
  \bibinfo{author}{Mohamed, A.r.}, \bibinfo{author}{Dahl, G.E.},
  \bibinfo{author}{Saon, G.}, \bibinfo{author}{Soltau, H.},
  \bibinfo{author}{Beran, T.}, \bibinfo{author}{Aravkin, A.Y.},
  \bibinfo{author}{Ramabhadran, B.}, \bibinfo{year}{2013}.
\newblock \bibinfo{title}{Improvements to deep convolutional neural networks
  for lvcsr}, in: \bibinfo{booktitle}{2013 IEEE workshop on automatic speech
  recognition and understanding}, \bibinfo{organization}{IEEE}. pp.
  \bibinfo{pages}{315--320}.
\bibitem[{Salimbajevs and
  Kapo{\v{c}}i{\=u}t{\.e}-Dzikien{\.e}(2022)}]{salimbajevs2022automatic}
\bibinfo{author}{Salimbajevs, A.},
  \bibinfo{author}{Kapo{\v{c}}i{\=u}t{\.e}-Dzikien{\.e}, J.},
  \bibinfo{year}{2022}.
\newblock \bibinfo{title}{Automatic speech recognition model adaptation to
  medical domain using untranscribed audio}, in: \bibinfo{booktitle}{Digital
  Business and Intelligent Systems: 15th International Baltic Conference,
  Baltic DB\&IS 2022, Riga, Latvia, July 4--6, 2022, Proceedings},
  \bibinfo{organization}{Springer}. pp. \bibinfo{pages}{65--79}.
\bibitem[{Schneider et~al.(2019)Schneider, Baevski, Collobert and
  Auli}]{schneider2019wav2vec}
\bibinfo{author}{Schneider, S.}, \bibinfo{author}{Baevski, A.},
  \bibinfo{author}{Collobert, R.}, \bibinfo{author}{Auli, M.},
  \bibinfo{year}{2019}.
\newblock \bibinfo{title}{wav2vec: Unsupervised pre-training for speech
  recognition}.
\newblock \bibinfo{journal}{arXiv preprint arXiv:1904.05862} .
\bibitem[{Seide et~al.(2011)Seide, Li and Yu}]{seide2011conversational}
\bibinfo{author}{Seide, F.}, \bibinfo{author}{Li, G.}, \bibinfo{author}{Yu,
  D.}, \bibinfo{year}{2011}.
\newblock \bibinfo{title}{Conversational speech transcription using
  context-dependent deep neural networks}, in: \bibinfo{booktitle}{Twelfth
  annual conference of the international speech communication association}.
\bibitem[{Shi et~al.(2020)Shi, Wang, Wu, Fuegen, Zhang, Le, Yeh and
  Seltzer}]{shi2020weak}
\bibinfo{author}{Shi, Y.}, \bibinfo{author}{Wang, Y.}, \bibinfo{author}{Wu,
  C.}, \bibinfo{author}{Fuegen, C.}, \bibinfo{author}{Zhang, F.},
  \bibinfo{author}{Le, D.}, \bibinfo{author}{Yeh, C.F.},
  \bibinfo{author}{Seltzer, M.L.}, \bibinfo{year}{2020}.
\newblock \bibinfo{title}{Weak-attention suppression for transformer based
  speech recognition}.
\newblock \bibinfo{journal}{arXiv preprint arXiv:2005.09137} .
\bibitem[{Su et~al.(2019)Su, Zhu, Cao, Li, Lu, Wei and Dai}]{su2019vl}
\bibinfo{author}{Su, W.}, \bibinfo{author}{Zhu, X.}, \bibinfo{author}{Cao, Y.},
  \bibinfo{author}{Li, B.}, \bibinfo{author}{Lu, L.}, \bibinfo{author}{Wei,
  F.}, \bibinfo{author}{Dai, J.}, \bibinfo{year}{2019}.
\newblock \bibinfo{title}{Vl-bert: Pre-training of generic visual-linguistic
  representations}.
\newblock \bibinfo{journal}{arXiv preprint arXiv:1908.08530} .
\bibitem[{Vedantam et~al.(2015)Vedantam, Lawrence~Zitnick and
  Parikh}]{vedantam2015cider}
\bibinfo{author}{Vedantam, R.}, \bibinfo{author}{Lawrence~Zitnick, C.},
  \bibinfo{author}{Parikh, D.}, \bibinfo{year}{2015}.
\newblock \bibinfo{title}{Cider: Consensus-based image description evaluation},
  in: \bibinfo{booktitle}{Proceedings of the IEEE conference on computer vision
  and pattern recognition}, pp. \bibinfo{pages}{4566--4575}.
\bibitem[{Yan and Pei(2022)}]{yan2022clinical}
\bibinfo{author}{Yan, B.}, \bibinfo{author}{Pei, M.}, \bibinfo{year}{2022}.
\newblock \bibinfo{title}{Clinical-bert: Vision-language pre-training for
  radiograph diagnosis and reports generation}, in:
  \bibinfo{booktitle}{Proceedings of the AAAI Conference on Artificial
  Intelligence}, pp. \bibinfo{pages}{2982--2990}.
\bibitem[{Yu et~al.(2022)Yu, Endo, Krishnan, Pan, Tsai, Reis, Fonseca, Lee,
  Abad, Ng et~al.}]{yu2022evaluating}
\bibinfo{author}{Yu, F.}, \bibinfo{author}{Endo, M.},
  \bibinfo{author}{Krishnan, R.}, \bibinfo{author}{Pan, I.},
  \bibinfo{author}{Tsai, A.}, \bibinfo{author}{Reis, E.P.},
  \bibinfo{author}{Fonseca, E.K.U.N.}, \bibinfo{author}{Lee, H.M.H.},
  \bibinfo{author}{Abad, Z.S.H.}, \bibinfo{author}{Ng, A.Y.}, et~al.,
  \bibinfo{year}{2022}.
\newblock \bibinfo{title}{Evaluating progress in automatic chest x-ray
  radiology report generation}.
\newblock \bibinfo{journal}{medRxiv} , \bibinfo{pages}{2022--08}.
\bibitem[{Zeghidour et~al.(2018)Zeghidour, Xu, Liptchinsky, Usunier, Synnaeve
  and Collobert}]{zeghidour2018fully}
\bibinfo{author}{Zeghidour, N.}, \bibinfo{author}{Xu, Q.},
  \bibinfo{author}{Liptchinsky, V.}, \bibinfo{author}{Usunier, N.},
  \bibinfo{author}{Synnaeve, G.}, \bibinfo{author}{Collobert, R.},
  \bibinfo{year}{2018}.
\newblock \bibinfo{title}{Fully convolutional speech recognition}.
\newblock \bibinfo{journal}{arXiv preprint arXiv:1812.06864} .

\end{thebibliography}

\end{document}